\documentclass{PoS}
\usepackage{amsmath}
\usepackage{wrapfig}

\title{Cosmic rays in early star-forming galaxies and their effects on the interstellar medium}

\ShortTitle{CRs in early star-forming galaxies}

\author{\speaker{Ellis R. Owen},$^{a, b}$ Kinwah Wu,$^{a, d}$ Pooja Surajbali$^{c}$ and Idunn B. Jacobsen$^{a}$\\
     \llap{$^{a}$}Mullard Space Science Laboratory, University College London, Holmbury St. Mary, Dorking, Surrey, RH5 6NT, United Kingdom \\
     \llap{$^{b}$}Department of Physics, National Tsing Hua University, Hsinchu, Taiwan (ROC)
     \llap{$^c$}Max-Planck-Institut f\"{u}r Kernphysik, Saupfercheckweg 1, Heidelberg 69117, Germany\\
          \llap{$^d$}School of Physics, University of Sydney, NSW 2006, Australia\\
     E-mails:  \email{ellis.owen.12@ucl.ac.uk}, \email{kinwah.wu@ucl.ac.uk}, \email{pooja.surajbali@mpi-hd.mpg.de}, \email{idunn.jacobsen.09@ucl.ac.uk}
      }

\abstract{Galaxies at high redshifts with strong star formation are sources of high-energy cosmic rays. These cosmic rays interact with the baryon and radiation fields of the galactic environment via photo-pair, photo-pion and proton-proton processes to produce charged and neutral pions, neutrons and protons. The cosmic rays thereby deposit energy into the interstellar medium (ISM) as they propagate. We show how energy transport and deposition by ultra high-energy cosmic rays is regulated by the evolution of the galaxy, in particular by the development of the galactic magnetic field. We show how the particle-driven energy deposition can influence the thermal evolution of the host and its surroundings. Using a parametric protogalaxy model, we calculate the heating effect on the ISM as the cosmic rays are increasingly confined by the magnetic evolution of the galaxy.}

\FullConference{35th International Cosmic Ray Conference --- ICRC2017\\
		10--20 July, 2017\\
		Bexco, Busan, Korea}

\begin{document}

\section{Introduction}

High-redshift star forming protogalaxies are thought to be abundant in ultra-high-energy (UHE) cosmic ray (CR) particles. The star formation rates (SFR) of such galaxies yield high rates of supernova (SN) explosions
  that provide ample extreme environments in compact objects 
  (e.g.\ fast spinning neutron stars and accreting black holes) and supernova remnants (SNR). These are considered likely regions in which CR acceleration may reach energies above 10$^{17}$ eV (see e.g. ~\cite{Kotera2011, Jacobsen2015, Berezinsky2006, Dar2008PhR}).

In this work we employ a generic spherical protogalaxy model of 10$^6$ stars, intended as a parametric mean. It is defined by a density field (of characteristic radius 1 kpc and peak density 10 cm$^{-3}$ following a Dehnen profile~\cite{Dehnen1993}), a magnetic field (described by the model introduced in~\cite{Schober2013}, but spatially following the density profile) and a radiation field which is dominated by the cosmic microwave background (CMB). We set our model to exhibit extreme starburst characteristics with a high SFR of 1000 ${\rm M}_{\odot} {\rm yr}^{-1}$. From this, a SN rate can be estimated. A typical SN explosion is set to have an energy of $10^{51}$ erg, of which a fraction of $\xi = 0.1$ is transferred into CR proton energy. We model primary CR production as a population of high-energy protons following a power-law distribution in energy above a GeV, extending up to 10$^{11}$ GeV with spectral index -2.1, similar to CR source regions in the Milky Way~\cite{Kotera2011, Ackermann807}. The CR luminosity is determined by the SFR (via the SN rate, assuming stars of mass > $8 {\rm M}_{\odot}$ yield SN explosions, being around 4\% of the stellar population if adopting a Salpeter IMF of index 2.35).

\section{Cosmic Ray Interactions}

More than 99\% of the energy of a typical CR population is contained in the high-energy component, above a GeV~\footnote{Note that these energies refer to the CR kinetic energy, not total energy.} (see, e.g.~\cite{Benhanbiles-Mezhoud2013}), so these are most likely to drive a heating effect. While high-energy CRs can deposit their energy by collisional energy transfer/ionization, this is relatively inefficient. A 0.5 GeV CR proton would only lose around 2.5\% of its energy by this mechanism over a Hubble time at $z=7$~\cite{Sunyaev2015}.  Above a threshold of around a GeV, pion-producing hadronic interactions with the radiation and baryon fields of the galaxy are a much more effective means by which CRs can deposit energy into their surroundings. 

The interaction of CR protons with the radiation field - ${\rm p}\gamma$ interactions - can proceed via two fundamental channels. These are: 
(1) a Bethe-Heitler pair production process~\cite{Bethe1934} proceeding as
${\rm A}\gamma \rightarrow {\rm A}' {l}^+ {l}^-$,
   where ${\rm A}$ and ${\rm A}'$ are nucleons (i.e. the incident CR before and after the interaction), 
   and ${l}^+$ and ${l}^-$ are charged leptons (at the energies of interest, around 90\% of the produced secondaries are electrons and positrons, with around 10\% being heavier counterparts); and
 (2) a photo-pion process proceeding as a resonant single pion production through the formation of $\Delta^+$ particles ($\rm{p} \gamma \rightarrow \Delta^{+}$). These $\Delta^+$ decay through two major channels, where charged and neutral pions are produced. These pions decay as
	   $ \rm{p} \pi^0 \rightarrow \rm{p}2\gamma$ and
		$\rm{n} \pi^+ \rightarrow \rm{n} \mu^+ \nu_{\mu}	$ where $\mu^+ \rightarrow \rm{e}^+ \nu_e \bar{\nu}_{\mu}$
(see \cite{Berezinsky1993}).
The branching ratios for the $(\Delta^{+}\rightarrow\pi^0)$ and $(\Delta^{+}\rightarrow\pi^+)$ channels 
   are 2/3 and 1/3 respectively. The direct production of pions is less dominant, occurring at a fraction of the rate of the above interaction, so is ignored in this analysis.
   CRs above 0.28 GeV also undergo losses by interactions with the baryon field of the galaxy. These pion-production losses proceed as ${\rm pp} \rightarrow \rm{p}  \Delta^{+~\;}$ and ${\rm pp} \rightarrow \rm{n}  \Delta^{++}$ with $\Delta^{+}$ and $\Delta^{++}$ baryons as the resonances~\cite{Almeida1968, 2008EPJA...35..317S}.

In these channels, the charged pion decay is largely responsible for the deposition of energy in the medium. This is because they mostly decay to produce leptons (which scatter strongly and cool quickly in an ISM environment) and neutrinos via a weak interaction ($ \pi^+	\rightarrow \mu^+ \nu_{\rm \mu} \rightarrow \rm{e}^+ \nu_e \bar{\nu}_{\rm \mu} \nu_{\rm \mu}\ $ and $ \pi^-	\rightarrow \mu^- \bar{\nu}_{\rm \mu} \rightarrow \rm{e}^- \bar{\nu}_e \nu_{\rm \mu} \bar{\nu}_{\rm \mu} $)
  on a timescale of $2.6\times 10^{-8}\;\!{\rm s}$.  The $\pi^0$ production leads to $\gamma$-ray emission in an electromagnetic process ($\pi^0	\rightarrow 2\gamma$)	 
  over shorter timescales of $8.5 \times 10^{-16}\;\!{\rm s}$.   

\subsection{Interaction Cross Sections and Path Lengths}

The energy loss driven by an interaction can be characterised in terms of its inelastic cross section which is specific to each interaction. We may express an interaction cross section as a function of the invariant normalised CR energy, $\epsilon_{\rm r} = \gamma_{\rm p} (1-\beta_{\rm p})\epsilon$. Here $\mu = \cos \theta$ for $\theta$ as the angle between the momentum vectors of the CR and target photon or proton, $\epsilon$ is the energy of the target photon or proton normalised to electron rest mass (being $h \nu/m_{\rm e} c^2$ and $m_{\rm p}/m_{\rm e}$ respectively) and $\gamma_p$ is the Lorentz factor of an incident CR proton, i.e. $E_{\rm CR} = \gamma_{\rm p} m_{\rm p} c^2$. The cross section for photo-pair production may be approximated as
\begin{equation}%
\label{approx-sigma}
\hat{\sigma}_{\rm \gamma e}(\epsilon_{\rm r})\approx %
\left\{	\frac{7}{6\pi}\alpha_{\rm f} \ln\left[\frac{\epsilon_{\rm r}}{k_{\rm \gamma e}}\right] \right\} \sigma_{\rm T}
\end{equation}%
  with $k_{\rm \gamma e}$ taking a value of $\approx 6.7$\footnote{Note that this approximation follows from the treatment outlined in~\cite{Stepney1983} - see also~\cite{Jost1950, Bethe1954, Blumenthal1970}}. Here $\alpha_{\rm f}$ is the fine structure constant and $\sigma_{\rm T}$ is the Thomson scattering cross section.
Without losing generality we fix $k_{\rm \gamma e} = 6.7$ in our calculations. In line with~\cite{Dermer2009, Atoyan2003}, the photo-pion cross section can be estimated as $\hat{\sigma}_{\rm \gamma \pi} \approx  \hat{\sigma}_{\rm \gamma \pi}^*\;\! {\cal H}(\epsilon_{\rm r} - \epsilon_{\rm th})$, 
  where ${\cal H}(...)$ is the Heaviside step function. 
In our calculations we adopt $ \hat{\sigma}_{\rm \gamma \pi}^* = 70~\rm{\mu b}$ and 
 $\epsilon_{\rm th} = 390$. We adopt the analytic cross section parameterisation prescribed by~\cite{Kafexhiu2014} for the pp interaction.

We can account for the relative importance of these various processes as a function of energy by characterising the CR losses in terms of the effective mean free path over which a CR proton loses its energy. Following~\cite{Protheroe1996APh, Dermer2009} for the photo-pair interaction, this path length is
\begin{equation}%
\label{eq:photopair_losses}%
r_{\rm \gamma e}(\gamma_{\rm p}) \approx %
\mathcal{R}_{\rm m} \frac{1125}{14}
	  \frac{\lambda_{\rm C}^3 \gamma_{\rm p}^3 b^{5/3}}{ \alpha_{\rm f} \sigma_{\rm T}\mathcal{F}_{\rm \gamma e}} \hspace{0.2cm} \ ,%
\end{equation}%
in Mpc, where $\lambda_{\rm C}$ is the Compton wavelength, $\mathcal{R}_{\rm m} = \frac{m_{\rm p} }{m_{\rm e}}$ is the ratio of proton to electron mass, $b =  m_{\rm e} {\rm c}^2/\gamma_p {\rm k}_{\rm B} T$ and 
\begin{equation}%
\mathcal{F}_{\rm \gamma e} = %
	\mathcal{C}(b) - \mathcal{D}(b)\ln\left[\frac{b k_{\rm \gamma e}}{1.03}\right] - \left(\frac{b k_{\rm \gamma e}}{2.62}\right)^{3/2}~\ln[1-e^{-b}] \ .%
\end{equation}%
The functions $\mathcal{C}(b)$,  $\mathcal{D}(b)$ and $\mathcal{E}(b)$ result from standard integrals, with $\mathcal{C}(b) = 0.74$ and $\mathcal{D}(b) = \Gamma(5/2)\zeta(5/2)$ when $b\ll1$, $\mathcal{C}(b) = b^{3/2}\ln(b)e^{-b}$ and $\mathcal{D}(b) = b^{3/2}e^{-b}$ when $b\gg 1$~\cite{Dermer2009}. The photo-pion interaction yields a path length of
\begin{equation}%
\label{eq:photopion_losses}
r_{\rm \gamma \pi}   \approx  \frac{\lambda_{\rm C}^3}{16\pi \;\! \hat{\sigma}_{\rm \gamma \pi}^*}   
    \left( \frac{m_{\rm e}{\rm c}^2}{k_{\rm B}T}\right)^3 {\rm e}^{k\eta}\left[\sum_{k=1}^\infty \left(\frac{\eta}{k^2} +\frac{1}{k^3} \right) \right]^{-1} \ , 
  \end{equation}%
where
 $\eta = \epsilon_{\rm th}m_{\rm e}{\rm c}^2  / 2\gamma_{\rm p} {\rm k}_{\rm B}T$. 

The pp interaction path length can be reduced to the classical definition of the mean free path of a scattered particle, $ r_{\rm p\pi} = (\hat{\sigma}_{\rm p\pi} n_{\rm p})^{-1}$, which is valid when the target baryon field is assumed to be at rest. This is reasonable given that the energy of the particles which make up the target baryon field are generally of much lower energy than the UHECRs.

We find that ${\rm p}\gamma$ interactions with the CMB dominate the CR losses due to the radiation fields. Overall, however, these are only important above energies of around 10$^{19}$ eV. Below this, pp interactions govern the majority of the energy deposition and subsequent heating process by the UHECRs.

\subsection{Cosmic Ray Propagation}
\label{sec:crtrans}

\begin{wrapfigure}{r}{0.6\textwidth}
	\includegraphics[width=0.62\columnwidth]{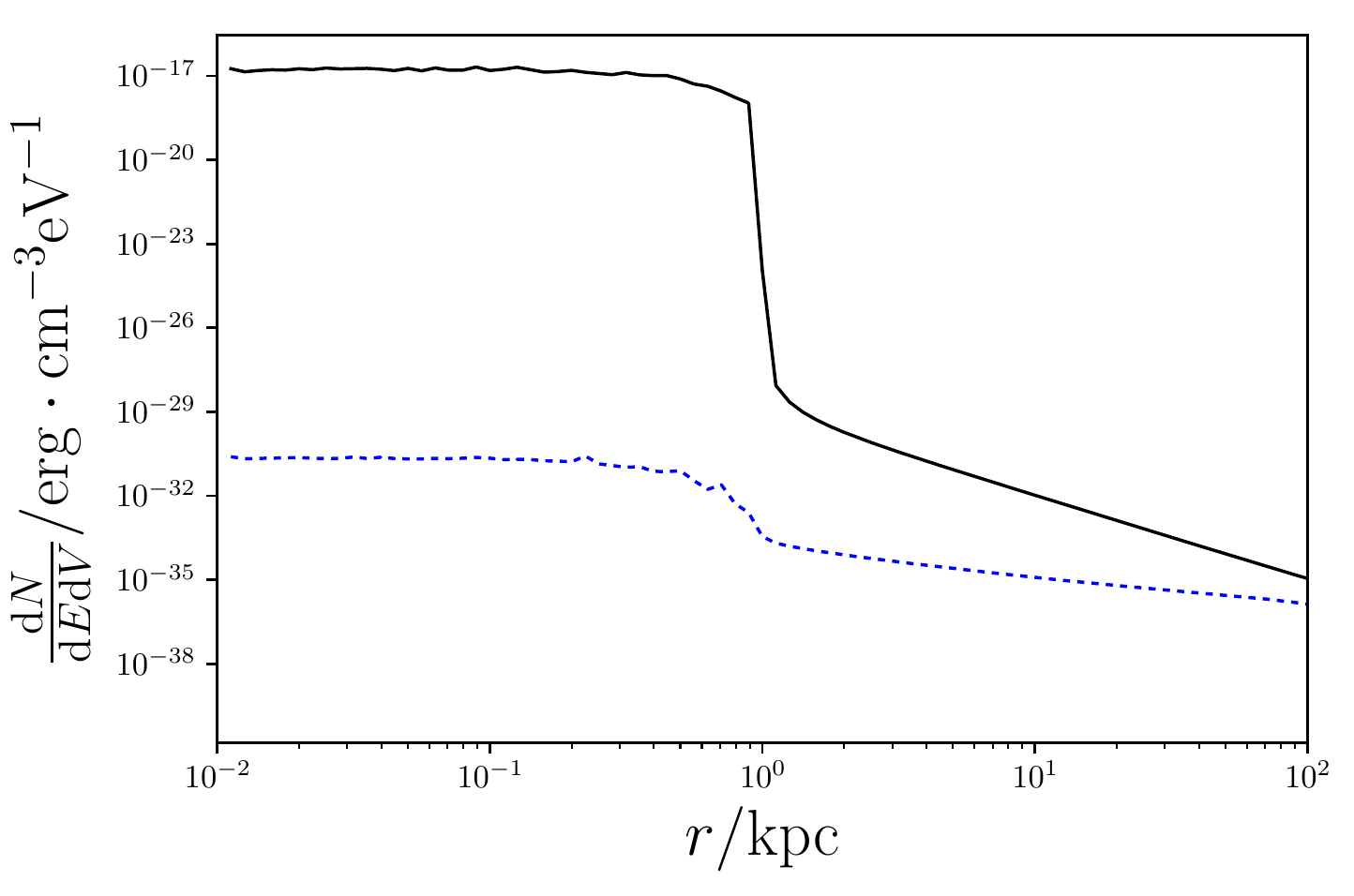}
    \caption{Diffusion profile after magnetic saturation (black). For comparison, we show the initial free-streaming case (blue).}
    \label{fig:diffusion_profile}
\end{wrapfigure}
High-energy CRs will travel at approximately the speed of light $c$ in free space, 
  but in the presence of a sufficiently strong magnetic field ($B\sim \mu{\rm G}$), 
  the propagation of CRs is practically a diffusive process, described by the transport equation: 
\begin{equation}
\label{eq:diffusion_equation}
\frac{\partial n}{\partial t} = \nabla \cdot \left[ D(E, r, t) \nabla n \right] + Q(r, E) \ ,
\end{equation} 
   where $n(E, r, t)$ is the local CR number density and $D(E, r, t)$ is the diffusion coefficient 
   (units $\text{cm}^2 ~ \text{s}^{-1}$).  
$Q(r, E)$ is a composite source term, consisting of an injection term and a loss term, 
  which are functions of CR energy, $E$. 
If the CR loss is insignificant, 
  the problem is essentially the transportation of a packet of CRs from the injection site across the system, i.e. the protogalaxy. 
Without losing generality, we may consider the system at first as an initial value problem (IVP) with $n(r, t=0) = Q = F_{\rm CR}(E) \cdot \mathcal{N}(r_s)$ erg s$^{-1}$ eV$^{-1}$, 
   where $r_s$ is the injection source location. The solution of this IVP is then integrated over time to yield the solution to the transport equation as 
\begin{equation}
\label{eq:diff_sol4}
N(E,r,t) = \frac{F_{\rm CR} \cdot \mathcal{N}(r_s) ~t}{\left(4\pi D(E, r, t) t \right)^{3/2}} \left\{\frac{(r-r_s)^2}{4 D(E, r, t) t}\right\}^{-{1}/{2}} \Gamma\left(\frac{1}{2}, \frac{(r-r_s)^2}{4  D(E, r, t) t}\right) + \mathcal{C}(E, r) 
\end{equation}
  (in units of erg cm$^{-3}$ eV$^{-1}$), 
  where $\mathcal{C}(E, r)$ is the constant of integration, 
  which can be derived from the initial freely streaming CR distribution.  
This solution corresponds to a continuous injection by a single source  
   (of a size $\sim 0.01\;\! {\rm pc}$ if it is a SNR, see, e.g.~\cite{Badenes2010}).  
A Monte-Carlo (MC) method may be used to sum over all the source contributions within the host galaxy distribution. 
In this work we weight the CR source contribution by the density profile of the galaxy. This is because we assume that the CR sources are remnants of stars, 
  which are ultimately derived from the gas profile from which stars form. 
Adopting a SFR of 1000 ${\rm M}_{\odot} {\rm yr}^{-1}$ and a magnetic field saturating at $10 \mu$G, 
  we obtain a CR profile as shown in Fig.~\ref{fig:diffusion_profile}. 
 
\subsection{Cooling Losses}

CR protons and their secondary electrons can lose energy in the ISM of high redshift galaxies through synchrotron, inverse Compton and free-free emission mechanisms. We can quantify the relative importance of a cooling mechanism by considering the respective CR energy loss timescale. At high energies and high redshifts, inverse Compton losses dominate from scattering off CMB photons~\cite{Schober2015, Schleicher2013}. The typical loss timescale as a function of secondary CR electron energy by this process is given by
\begin{equation}
\label{eq:losstime}
\tau(E) = \frac{3 m_{\rm e}^2 c^3}{4 \sigma_{\rm T} U_{\rm CMB} E}
\end{equation}
where $c$ is the speed of light, $m_{\rm e}$ is the mass of an electron and $U_{\rm CMB}$ is the energy density of the CMB,
\begin{equation}
U_{\rm CMB} = \frac{8 \pi^5 k_{\rm B}^4}{15 c^3 h^3}T_0^4(1+z)^4 \ .
\end{equation}
Here $h$ is the Planck constant, $T_0=2.73~{\rm K}$ is the current CMB temperature \cite{2015arXiv150201589P}, and $z$ is the redshift of interest - in this study, we take $z=7$.

\subsection{Sunyaev-Zel'dovich Effect}

Given the dominance of inverse Compton cooling of the CRs at high redshift, we may consider the impact of the resulting emission from these losses. The scattering of high-energy electrons and protons off CMB photons is known as the Sunyaev-Zel'dovich effect (SZE)~\cite{Sunyaev1972}. In general, the inverse-Compton power per unit volume per energy is given by
\begin{equation}
\frac{{\rm d} P(\epsilon)}{{\rm d}\epsilon} = \frac{{\rm d} E}{{\rm d} V {\rm d} t {\rm d}\epsilon} = \frac{N_{\rm e} 8 \pi^2 r_0^2}{h^3 c^2}\left[k_{\rm B} T_0 (1+z)\right]^{{(p+5)}/2}A(p) \Gamma\left(\frac{p+5}{2}\right)\zeta\left(\frac{p+5}{2}\right) \epsilon^{-(p-1)/2} \ ,
\end{equation}
in erg cm$^{-3}$ s$^{-1}$ eV$^{-1}$\cite{BlumenthalGould, Rybicki1979}, where $r_0$ is the classical electron radius, $p$ is the negative power law index of the CR spectrum and 
\begin{equation}
A(p) = 2^{p+3} \frac{p^2 + 4p + 11}{(p+3)^2 (p+5) (p+1)} \ .
\end{equation} 
From equation~\ref{eq:losstime}, it can be seen that the SZ emission due to electrons dominates that by protons by around three orders of magnitude (due to their mass ratio). While we have assumed that primary CR electrons deposit their energy relatively close to their source (and so have ignored their contribution in the present work), we can account for the SZ emission due to the secondary CR electrons produced from the interaction showers of the primary CR protons.

We employ a secondary CR electron injection model based on the injection of primary CR protons $Q_{\rm p}(\gamma_{\rm p})$, as in~\cite{Schober2015, Lacki2013}. This gives a CR electron injection rate of
\begin{equation}
\label{eq:electronproton}
Q_{\rm e}(\gamma_{\rm e}) \simeq \frac{\Upsilon}{6}\frac{400~m_{\rm e}}{m_{\rm p}} Q_{\rm p}(\gamma_{\rm p}) \ ,
\end{equation}
which is dependent on the energy of the primary CR by $\gamma_{\rm p} = E/m_{\rm p} c^2$ and where $\gamma_{\rm e} = E/m_{\rm e} c^2$. $\Upsilon$ is the fraction of energy of the original CR which is transferred to the electron and positron producing branches of the particle shower, taken to be 0.5~\cite{Lacki2011}. $N_{\rm e}$ can be estimated from $Q_{\rm e}*\tau_{\rm cool}*\mathcal{X}(r)$.

We can estimate the X-ray luminosity between $E_{\rm min} = 100$ eV and $E_{\rm max} = 100$ keV by considering the inverse-Compton energy that goes into this band. We take the emission is this band as the source of X-ray heating. The power emitted in this band per unit volume is
\begin{equation}
P(\epsilon) = \frac{N_{\rm e} 8 \pi^2 r_0^2}{h^3 c^2}\left[k_{\rm B} T_0 (1+z)\right]^{{(p+5)}/2}B(p)~\Gamma\left(\frac{p+5}{2}\right)\zeta\left(\frac{p+5}{2}\right) \left[E_{\rm max}^{(3-p)/2}-E_{\rm min}^{(3-p)/2}\right] \ ,
\end{equation}
where $B(p) = 2 A(p)/(3-p)$ and with the constraint that $p \neq 3$. For comparison, this indicates a total SZE X-ray luminosity of the star forming galaxy of up to around 10$^{48}$ erg s$^{-1}$ in the 0.5 to 8 keV band often used in observational studies. Local and distant starbursts exhibit X-ray luminosities of around 5 orders of magnitude lower than this (see~\cite{Wang2013}), perhaps indicative of the role played by heating-driven outflows and other CR escape mechanisms to reduce the true CR containment level. This more detailed modelling is left to future work.

\section{Heating Processes}

\begin{wrapfigure}{r}{0.6\textwidth}
	\includegraphics[width=0.62\columnwidth]{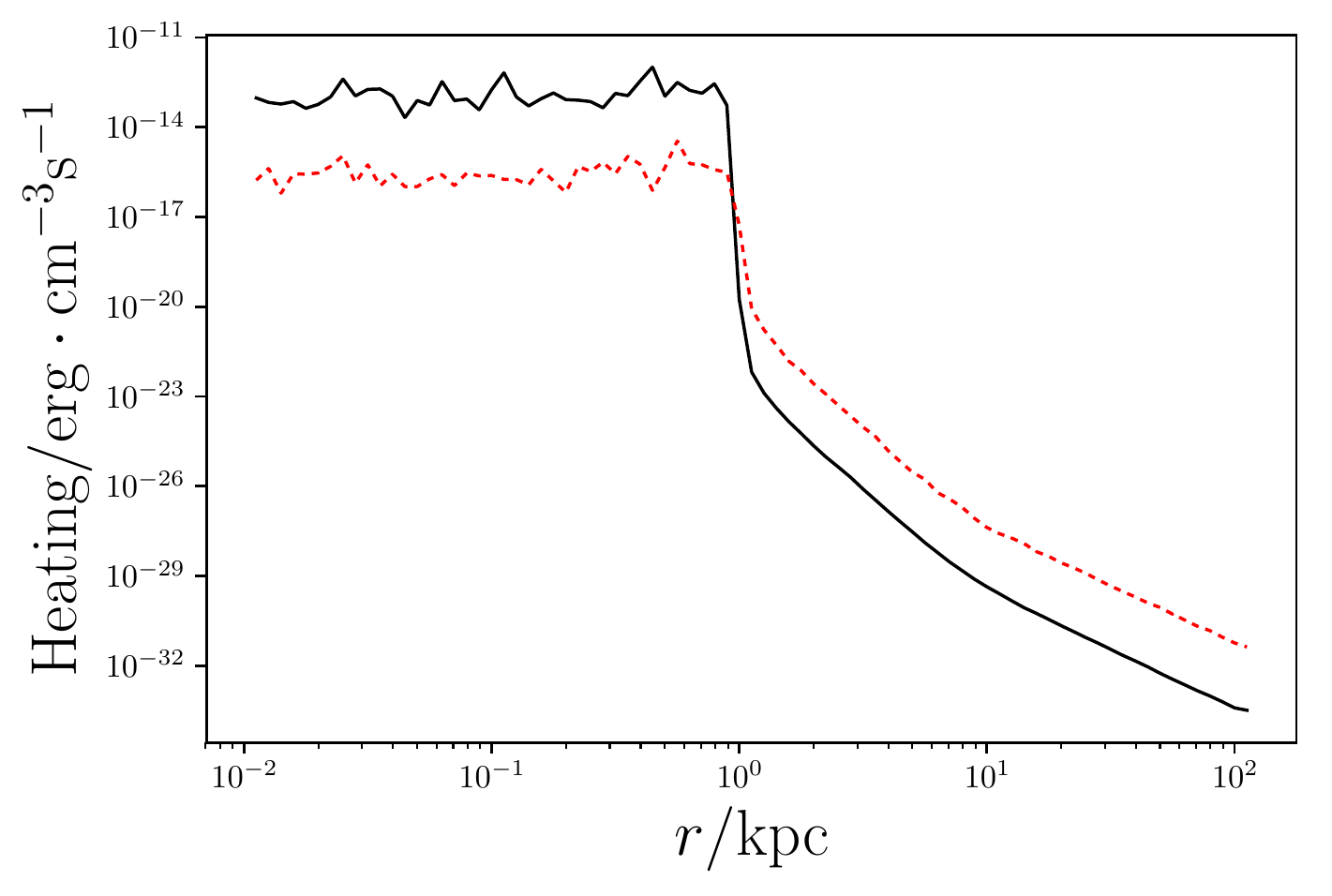}
    \caption{Heating due to CRs (black) and X-rays (red dashed) through the model}
    \vspace{-1em}
    \label{fig:heating_profile}
\end{wrapfigure}
The heating at a position $r$ in a medium irradiated by X-rays or CRs is calculated as the tendency of the medium to absorb radiation at that point (given by the absorption coefficient $\alpha(r) = n(r)~\sigma$) accounting for the attenuation along the particle or radiation paths. For the source distribution in the protogalaxy model considered here, we invoke an MC method in which the total source luminosity is split across an ensemble of points. Additionally, we introduce a scaling parameter, $\mathcal{X}(r)$, to account for the contained profile of CRs due to the galactic magnetic field. This is the ratio between the saturated diffusion result and the free-streaming result outlined in section~\ref{sec:crtrans} as a function of position. For CR heating, the absorption parameter is defined with the cross section for the pp interaction used. For X-ray heating, the same approach may be used but here the Klein-Nishina cross section is applied instead. The results are shown in Fig.~\ref{fig:heating_profile} for the heating profile due to the CRs (black), and X-rays (red).

\section{Summary and Conclusions}

Our calculations showed that an ISM CR heating rate may reach $10^{-13}$ erg cm$^{-3}$ s$^{-1}$ 
  and an indirect X-ray heating effect via the SZE of 10$^{-16}$ erg cm$^{-3}$ s$^{-1}$.  
As CR losses have not been accounted for explicitly in the transportation, these values may be considered as upper limits.  
Nevertheless, CR heating should not be ignored in protogalactic environments.  
Such heating would increase the Jeans mass of star-forming regions, 
   thus distorting the initial mass function of early generations of stars. 
Also, the resulting X-rays may increase the ionisation fraction deep in star forming clouds, 
  hampering the removal of magnetic fields by ambipolar diffusion~\cite{Draine2011}. 
Coupled with CR heating, which is expected to be particularly concentrated along the strong magnetic field vectors in star-forming regions, an amplified run-away heating effect may take hold, 
  resulting in quenching the star formation in some regions entirely.

\section*{Acknowledgements}

ERO is supported by the Science and Technology Facilities Council, a Royal Astronomical Society grant, the Institute of Physics (IoP) C R Barber Trust and the IoP Student Conference Fund.

\bibliographystyle{JHEP}
\bibliography{references}

\end{document}